# Electrically driven robust tuning of lattice thermal conductivity

E Zhou[1†], Donghai Wei[1†], Jing Wu[1], Guangzhao Qin,[1*] and Ming Hu[2*]

[1]*State Key Laboratory of Advanced Design and Manufacturing for Vehicle Body, College of Mechanical and Vehicle Engineering, Hunan University, Changsha 410082, P. R. China*
[2] *Department of Mechanical Engineering, University of South Carolina, Columbia, SC 29208, USA*

[†] These authors contributed equally.

[*] Corresponding authors: G.Q. <gzqin@hnu.edu.cn>, M.H. <hu@sc.edu>

**Abstract**

Two-dimensional (2D) materials represented by graphene stand out in future electrical industry and have been widely studied. As a commonly existing factor in electronic devices, the electric field has been extensively utilized to modulate the performance. However, how the electric field regulates thermal transport is rarely studied. Herein, we investigate the modulation of thermal transport properties by applying the external electric field ranging from 0 to 0.4 VÅ$^{-1}$, with bilayer graphene, monolayer silicene, and germanene as study cases. The monotonic decreasing trend of thermal conductivity of all the three materials is revealed. The significant effect on the scattering rate is found to be responsible for the decreased thermal conductivity by electric field. Further evidences show that the reconstruction of internal electric field and the generation of induced charges lead to the increased scattering rate from strong phonon anharmonicity. Thus, the ultra-low thermal conductivity emerges with external electric field applied. Applying external electric field to regulate thermal conductivity enlightens the constructive idea for high-efficient thermal management.

# 1. Introduction

In 2004, Geim *et al.* found that graphene could stably exist in the environment by 'peeling' method to exfoliation.[1] Inspired by graphene, more two-dimensional (2D) materials such as the IV group analogues[2,3], the V group analogues[4,5], MXene[6–8], transition metal dichalcogenides[9], and BiXene[10] emerge and have been widely explored for the outstanding physical and chemical properties. The applications of the 2D materials have promoted the development of thermoelectricity, photoelectricity, nanoelectronics, *etc*. However, the currently reported materials with excellent properties are still powerless for practical applications when facing with the ever-increasing requirements in extreme conditions. To purposefully obtain better properties, lots of effective control strategies like strain engineering, patterned cuts, doping, defects, rotating between layers, external magnetic field, and external electric field have been adopted.[4,11–26] These superior strategies greatly expand the applications of the pristine materials by leading to extraordinary properties.

Typically, materials with high thermal conductivity ($\kappa$) are mainly applied for heat dissipation, which have extensive applications in lots of fields, particularly in large CPU cluster, space vehicle, and nuclear reactor[27]. Unfortunately, the semimetal electronic nature of graphene prohibits its potential applications despite the ultrahigh thermal conductivity. From 1994, silicene, the counterpart of silicon to graphene, is well proven to have a promising future in the field of electronics from a theoretical perspective. However, the air stability of silicene limits the practical application.[28–30] In 2015, Li *et al*. achieved the encapsulation of silicene by 'silicene encapsulated delamination with native electrodes' (SEDNE) method, which can be also applied in fabricating germanene.[31] Besides, they reported a new type of silicene based field effect transistor (FET) and verified its better electronic characteristics compared with graphene, which consists with theoretical predictions, such as a small bandgap and high carrier mobility. Thus, 2D materials represented by silicene have aroused a research upsurge in the field of nanoelectronics. As a major factor affecting the performance and life of electronic devices, the heat dissipation ability quantified by thermal conductivity is a key parameter for device design. However, despite the common existence of electric field in electronics, the effect of electric field on thermal transport is rarely studied before. Note that there is no influence of external electric field on the thermal conductivity of monolayer graphene, which is restricted by the symmetry.[32] As for bilayer silicene and germanene, there are covalent bonds between layers.[33,34]

In this paper, the external electric field is employed to explore the influence on the thermal conductivity of the AB stacked bilayer graphene[35], monolayer silicene, and germanene[36]. We demonstrate that the thermal conductivity is negatively correlated with the applied electric field, which decreases about 1-3 orders of magnitude for the bilayer graphene, monolayer silicene and germanene. A series of fundamental insight from the scattering rate, dielectric constant, and microscopic view of electronic structure is conducted in the following parts to find out the inherent mechanism for the electrically driven robust tuning of lattice thermal conductivity. The significantly modulated thermal conductivity of the bilayer graphene, silicene, and germanene not only verifies the possibility of using electric field to tune thermal transportation, but also provides a remarkable scheme for the applications in other more fields.

## 2. Computational Methodology

The first-principles calculation is applied based on the density functional theory (DFT), choosing Perdew-Burke-Ernzerhof of generalized gradient approximation (PBE-GGA)[37] as exchange-correlation functional, by conducting the Vienna *ab initio* simulation package (VASP)[38], following by solving the phonon Boltzmann transport equation (BTE). The kinetic energy cutoff, energy convergence threshold, and the Hellmann-Feynman force are taken as 1000 eV, $10^{-8}$ eV, and $10^{-8}$ eVÅ$^{-1}$, respectively. The maximum strength of the electric field applied in the *out-of-plane* direction is up to 0.4 VÅ$^{-1}$ to make sure all the structures hold stability, where there is no imaginary frequency in the phonon spectrum. Noting that the dipole correction is taken into account throughout the calculation process, and the *Van der Waals* (vdW) forces is applied for the bilayer graphene in AB stacking. The 15×15×1 Monkhorst-Pack[39] *k*-point grid was used to sample the first Brillouin zone, and a vacuum layer of 20 Å was employed along the *out-of-plane* direction to avoid layer-to-layer interaction caused by periodic boundary conditions.

Based on the density functional perturbation theory (DFPT), the born effective charge and dielectric constant ($\varepsilon$) were calculated. Using the real-space finite displacement difference method applied in Phonopy[40] and thirdorder.py[41], the harmonic and non-harmonic force constants are obtained, respectively. The scattering matrix can be constructed based on the third-order force constant, and then we can calculate the phonon scattering rate and get the phonon relaxation time. Based on the convergence test, the 5th nearest neighbor is chosen as the cutoff radius for achieving the converged thermal conductivity. The $\kappa$ is obtained by using the ShengBTE software package[42] through solving the phonon BTE iteratively.

## 3. Results

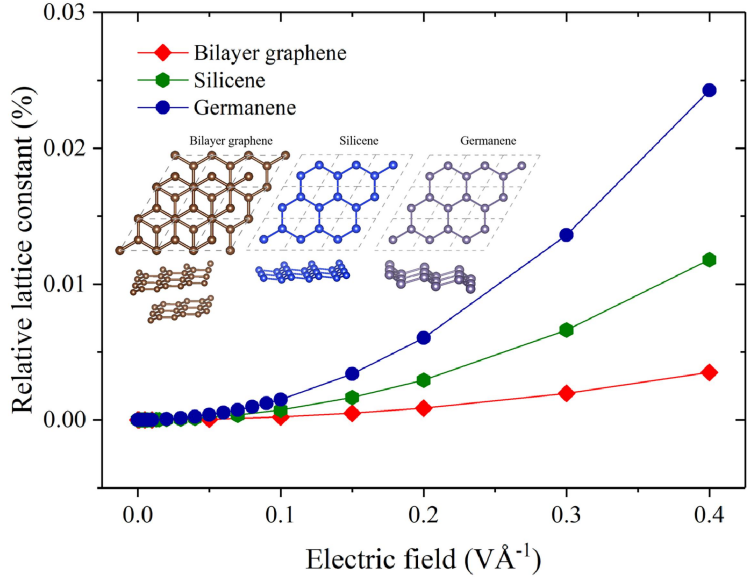

Fig. 1. The variation of lattice constant of AB stacking bilayer graphene, silicene, and germanene with respect to the strength of external electric field ($E_z$) along the *out-of-plane* direction, where the relative lattice constant is calculated as $\left(L_{E_z} - L_{(E_z=0)}\right) / L_{(E_z=0)}$ (Inset: the top and side views of structures of the optimized AB-stacked bilayer graphene, monolayer silicene, and germanene).

Different from the planar honeycomb structure of graphene, both silicene and germanene have non-planar bucking structure, as shown in the insert of Fig. 1. The thickness considering vdW distance of germanene (4.889 Å)[43] are larger than that of silicene (4.650 Å)[44], which may be due to its larger relative atomic mass and radius. The lattice constants of the three materials increase exponentially with the increasing electric field strength in the *z*-direction (perpendicular to the plane). Despite the exponential increase of lattice constant, the variation of the absolute value is tiny (<0.03%). Moreover, the structural characteristics of the three materials are slightly affected and the structure keeps buckling for silicene and germanene. In addition, it is worth mentioning that the flatter the structure, the less effect of the electric field on the lattice constant. On the contrary, the larger the bucking distance is, the more obvious lattice constant changes with the electric field applied.

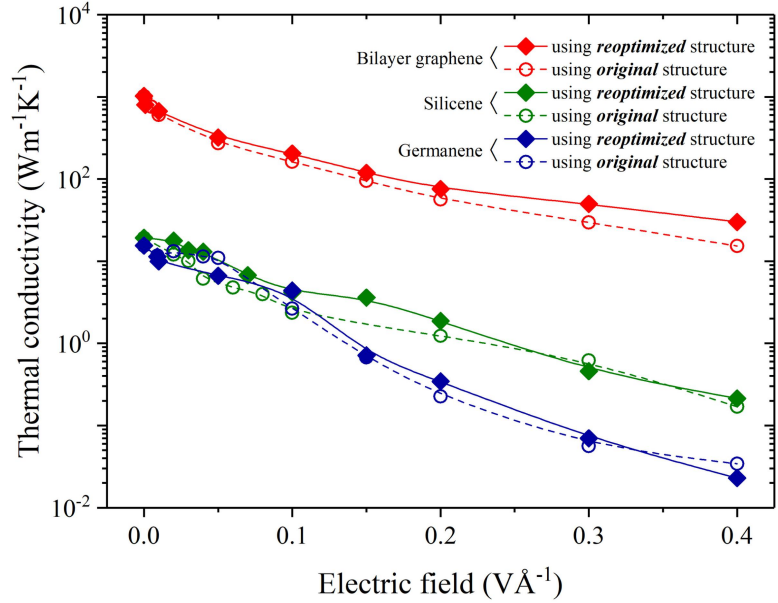

Fig. 2. The lattice thermal conductivity of bilayer graphene, monolayer silicene and germanene with the applied electric field ranging from 0 to 0.4 VÅ$^{-1}$. The original and reoptimized structures correspond to the calculations based on the pristine structure without electric field applied and the fully optimized structure with electric field applied, respectively.

To examine the effect of the electric field on the thermal transport properties of materials, we calculated the $\kappa$ of bilayer graphene, monolayer silicene, and germanene with electric field applied. The variation of $\kappa$ as a function of the strength of the electric field are presented in Fig. 2. Generally, the $\kappa$ of monolayer silicene and germanene are 19.21 and 15.50 Wm$^{-1}$K$^{-1}$, respectively, much lower than that of bilayer graphene (1021.86 Wm$^{-1}$K$^{-1}$) obtained with no external electric field, which are in good agreement with previous reports.[3,45,46] With the increasing external electric field, the $\kappa$ of the three materials shows the similar decreasing trend. Surprisingly, when $E_z$ = 0.4 VÅ$^{-1}$, the $\kappa$ of germanene decreases largely to 0.034 Wm$^{-1}$K$^{-1}$, which is nearly three orders of magnitude lower than that of the original $\kappa$ and is close to that of air.

In addition, considering the tiny effect of electric field on the geometry parameters, we further calculated the $\kappa$ based on the original structure for making a comparison, where the pristine structure without electric field applied is used for all the calculations with electric field applied. Noticeably that there is almost no difference of the $\kappa$ of the three materials when the structure is optimized or not. The results indicate that the geometry variation is not responsible for the giant effect of electric field on $\kappa$, and there must be other factors dominate the variation of $\kappa$.

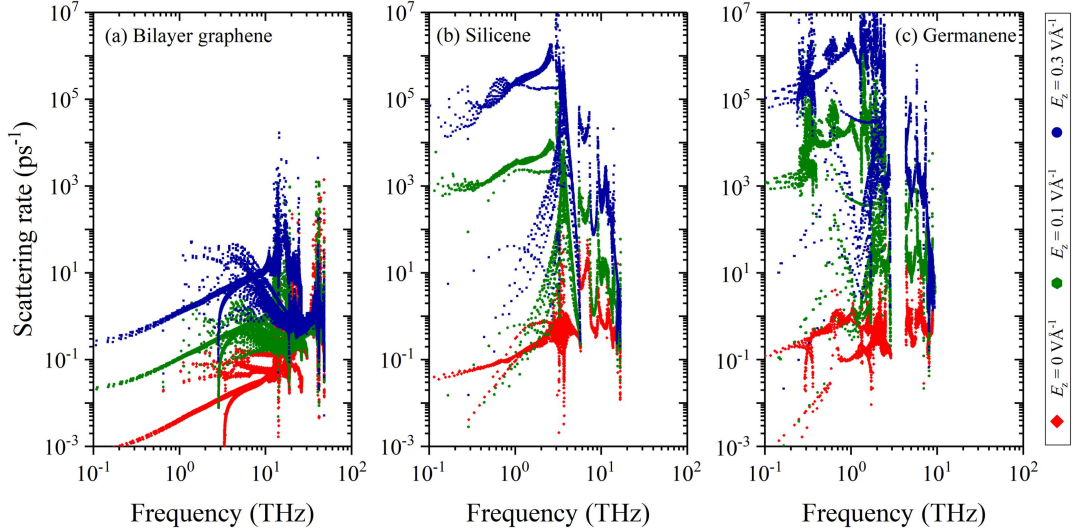

Fig. 3. The phonon scattering rate under the representative electric field (red for 0 VÅ$^{-1}$, green for 0.1 VÅ$^{-1}$, blue for 0.3 VÅ$^{-1}$) of the (a) bilayer graphene, (b) monolayer silicene, and (c) germanene.

It has been reported that the external electric field has very slight effect on the phonon spectrum[47], which is also confirmed in this study for bilayer graphene, monolayer silicene and germanene. Consequently, group velocities and specific heat capacity of bilayer graphene, monolayer silicene, germanene are almost not affected by the electric field. Thus, the only left factor is the relaxation time, which must be the primary reason for the decreasing and unique responding trends of $\kappa$[24,48] as revealed in Fig. 2. To figure out the issue, we accessed the variation of the scattering rate with respect to the electric field strength, as schematized in Fig. 3. When the strength of $E_z$ keeps increasing, the scattering rate of the three materials increases significantly. Additionally, the scattering intensity of bilayer graphene is much lower than those of monolayer silicene and germanene under the same external electric field such as $E_z$ = 0.3 VÅ$^{-1}$, which explains why the $\kappa$ of bilayer graphene is much higher than that of silicene and germanene.

Besides, germanene has the highest scattering rate, especially in the region of large electric field, which shows a variation trend consistent with the $\kappa$ discussed above. It can be clearly seen that the scattering rate of bilayer graphene varies less than those of monolayer silicene and germanene, which means that the reduction rate of the $\kappa$ is lower for bilayer graphene. When a small electric field is applied, although it can be found that there is a step change of the scattering rate for low frequency acoustic phonon branches of silicene and germanene, the ZA branch does not dominate the $\kappa$ of the two materials, which leads to the soft decease of $\kappa$ as shown in Fig. 2.

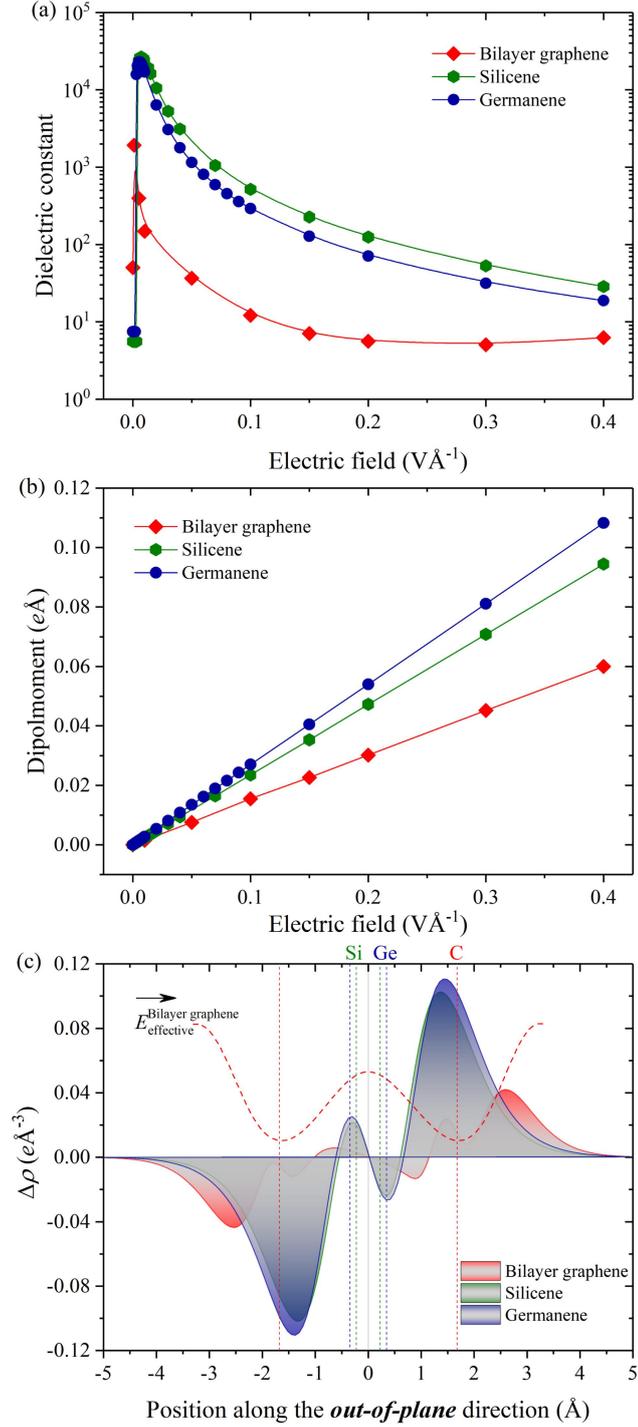

Fig. 4. The comparison for the bilayer graphene, monolayer silicene and germanene of (a) the dielectric constant and (b) the dipole moment under a few typical external electric fields. (c) The electric field induced charge redistribution with respect to the position along the *out-of-plane* plane, where $\Delta\rho = \rho_{(E_Z=0.1)} - \rho_{(E_Z=0)}$.

To explore the intrinsic mechanism for the electric field regulating thermal conductivity, we further investigate the variation from the perspective of microscopic electronic structure. The dielectric constant (Fig. 4a) and the dipole moment (Fig. 4b) with the $E_z$ ranging from 0 to 0.4 VÅ$^{-1}$ are extracted, as well as the charge density variation (Fig. 4c) along the $z$ direction as $E_z$ = 0.1 VÅ$^{-1}$ of bilayer graphene, silicene and germanene are shown. When $E_z > 0$ (Fig. 4a, b), the internal electric field is destroyed, resulting in the rise of the transfer of charges and the electrical polarization of the three materials. The dielectric constant of the three materials has a step change that reaches value four orders of magnitude larger, in particular, for monolayer silicene and germanene. As revealed in previous studies, the dielectric constant decays exponentially with the increasing electric field strength, indicating their weaker ability to restrain charges.[24] According to an accepted rule of thumb, the dipole moment is determined by the distance between the centers of positive and negative charges and their amounts. It can be seen that the dipole moment of the three materials increases monotonously with respect to electric field, which is derived from the small effect of electric field on the thickness of the structure (Fig. 1), and more charges will be generated in the meantime.

Evidently, there is a similar phenomenon for the charge density distribution (Fig. 4c) of silicene and germanene, which have two peaks with the similar values along the positive and negative directions of the $z$-axis, respectively. The situation is fundamentally different from bilayer graphene.[4,25,49] Taking the positive direction of the $z$-axis as an example, there are a positive charge accumulation peak and a negative charge formed sub-peak, which are caused by internal charge transfer and induced charge, respectively. The fundamental reason is the destruction of built-in field and then rebuilt, which can be traced back to the breaking inversion symmetry of structure when external electric field applied. Thus, the positive charge moves along with the positive direction of the $z$-axis, which is identical to the direction of electric field applied, while the negative charge moves oppositely. The charge transfer caused by the external electric field leads to the generation of internal opposite induced charges (located around the atoms), resulting in a reverse peak with opposite charges. It is intriguing to find that the peaks of induced charges of monolayer silicene and germanene both appear at the interface of the structure. In contrast, for the bilayer graphene, due to the vacuum layer between the two layers of graphene, the induced charge is deviated, resulting in the unique performance of the charge density and internal effective electric field. Prominently, electron redistribution caused by electric field applied will eventually increase phonon-phonon scattering, phonon anharmonicity, and further reduce the $\kappa$ of AB stacked bilayer graphene, monolayer silicene, and germanene.

## 4. Conclusion

In summary, the regulation on the thermal transport properties for bilayer graphene, monolayer silicene and germanene are studied by applying external electric field ranging from 0 to 0.4 VÅ$^{-1}$. The results reveal that the $\kappa$ of the three materials can be effectively modulated by the external electric field. As the electric field strength increases, the $\kappa$ decreases significantly. Especially, the $\kappa$ of monolayer germanene can even drop to a record low value of 0.034 Wm$^{-1}$K$^{-1}$ when $E_z$ reaches 0.4 VÅ$^{-1}$, which mainly results from the increased phonon-phonon scattering intensity. Combined with the microscopic electronic structure chart, the reconstruction of the internal electric field and the generation of induced charges are the essential reasons for the decrease of $\kappa$, which leads to the re-normalization of the interaction between atoms. Eventually, phonon-phonon scattering and phonon anharmonicity are affected by the renormalized interatomic interactions, which further leads to the decrease of the $\kappa$. The strategy by conducting the external electric field to manipulate $\kappa$ as shown in this paper provides a new insight to design and develop materials with great practical application value. It can be expected to have potential applications in nanoelectronics, thermoelectrics, thermal management, *etc*.

## Acknowledgements

This work is supported by the National Natural Science Foundation of China (Grant No. 52006057), the Fundamental Research Funds for the Central Universities (Grant Nos. 531119200237 and 541109010001), the Changsha Municipal Natural Science Foundation (Grant No. kq2014034), the State Key Laboratory of Advanced Design and Manufacturing for Vehicle Body at Hunan University (Grant No. 52175011), and the NSF (award number 2030128). Research reported in this publication was also supported in part by the NSF (award number 2030128). The numerical calculations in this paper have been done on the supercomputing system of the National Supercomputing Center in Changsha.